\documentclass[aps,onecolumn,10pt]{revtex4}
\usepackage[utf8]{inputenc}
\usepackage{amsmath,amssymb}

\usepackage{graphicx}
\usepackage{hyperref}
\usepackage{epstopdf}
\usepackage{physics}
\usepackage{subcaption}     % for subfigures with captions
\usepackage{caption}
\begin{document}

\title{Thermodynamics of BTZ Black Holes in Bumblebee Gravity with Barrow Entropy with Cavity-Modification }

\author{H. Kaur${}^{a}$}

\author{Prince A Ganai${}^{a}$}

\affiliation  {Department of Physics , National Institute of Technology , Srinagar , Kashmir-190006 , India }

\begin{abstract}

We investigate the thermodynamics of $(2+1)$-dimensional Bañados–Teitelboim–Zanelli (BTZ) black holes in the Einstein–bumblebee gravity theory with spontaneous breaking of Lorentz symmetry caused by a nonzero vacuum expectation value of the vector field. We analyse corrections to black hole thermodynamics, including the non-extensive Barrow entropy, parametrising quantum-gravitational corrections to the Bekenstein–Hawking area law. We introduce the York cavity formalism by placing the black hole in a finite isothermal cavity to obtain a properly defined canonical ensemble. Here we arrive at the corrected temperature, free energy, and stability conditions of the BTZ black hole, demonstrating the interplay among the Lorentz-violating effects, the Barrow corrections to Entropy, and the boundary conditions within the cavity. Our results indicate that the collective impact of bumblebee dynamics and Barrow entropy significantly alters the phase structure, equilibrium configurations, and thermal stability of BTZ black holes. These findings provide insight into the implications of Lorentz violation and generalized entropy frameworks in lower-dimensional quantum gravity.

\end{abstract}

\maketitle

\section{Introduction}

Black hole thermodynamics is one of the pillars of modern theoretical physics, providing profound insights into the interplay of gravity, quantum mechanics, and statistical thermodynamics~\cite{Hawking1975, Wald2001, Bekenstein1973}. Among all the numerous black hole solutions, the BTZ black hole in $(2+1)$ dimensions occupies a privileged position due to its simplicity and richness of structure~\cite{BTZ1992, BHTZ1993, Carlip1995, Teitelboim1998, Quevedo2009}. The BTZ solution clearly shows how black hole dynamics and thermodynamics arise even in lower-dimensional spacetimes where gravitational degrees of freedom are severely truncated~\cite{Vargas2004, Banados1993}.

BTZ black holes possess event horizons, mass, angular momentum, and charge in $(2+1)$-dimensional anti-de Sitter (AdS) spacetime~\cite{Carlip1995, Dehghani2023}. Their Entropy, temperature, and other thermodynamic properties replicate the already familiar behaviour in higher dimensions. However, the lower-dimensional models use more precise, analytic, and numerical approaches. Therefore, the BTZ solution is a treasure trove of practical computations in quantum gravity and holography, especially in AdS/CFT correspondence~\cite{Banados1993, Lemos1996}.

One of the gravitation frontiers is the study of spontaneous Lorentz symmetry breaking, perhaps due to quantum gravity or Standard Model extensions~\cite{KosteleckySamuel1989, KosteleckyLehnert2001}. Bumblebee gravity with a new vector field, the bumblebee field, that acquires a nonzero vacuum expectation value, is a standard model for such symmetry violations~\cite{MalufNeves2021, Xia2022}. The incorporation of the bumblebee field modifies both the geometry and dynamics of black holes, with observable consequences on horizons, thermodynamic characteristics, quasinormal modes, and stability~\cite{Afrin2024, Islam2024, Mai2023, Chen2025, Chengjia2023, Chen2022, Pan2022, Jing2021}.

Recent work has shown that black holes within Einstein–Bumblebee gravity exhibit nonstandard properties, including violations of the first law of black hole mechanics, novel phase transitions, and altered Hawking radiation and shadows~\cite{Mai2023, Chen2025, Chengjia2023, Chen2022, Pan2022, Jing2021}. In particular, rotating and charged BTZ-like solutions have been explicitly shown within the bumblebee theory, validating the significance of this theory within lower-dimensional settings~\cite{Dehghani2023, Chengjia2023}.

Parallel research in quantum gravity has driven the Bekenstein–Hawking area law generalizations of black hole entropy, trying to embrace horizon quantum fluctuations and Planck-scale features~\cite{Bekenstein1973, Hawking1975}. An essential proposal towards this is the Barrow entropy, encompassing horizon fractalization due to quantum gravitational effects~\cite{Barrow2020, Barrow2021}. Barrow's approach modifies the classical formula for Entropy as follows:

\[
S_{\rm B} = \left(\frac{A}{4}\right)^{1+\Delta},
\]
where $A$ is the horizon area and $0 \leq \Delta \leq 1$ is the fractal deformation parameter. This entropy correction influences temperature, heat capacities, and phase transitions, and has opened a flood of work examining its effects in other black hole spacetimes~\cite{Majhi2023, Houndjo2022, Lymperis2021, Wang2021, Saridakis2021, Sheykhi2012, Ghorui2024}.
Black hole thermodynamics beyond the microcanonical ensemble requires equilibrium creation on a finite boundary in spacetimes with negative specific heat. York's cavity formalism is a strict approach: placing a black hole in a thermal cavity with fixed radius $R$ allows for canonical thermodynamic quantities and the study of phase transitions and equilibrium~\cite{York1986, BrownYork1989, Wald1993, Ganai2025, Ganaie2025, Islam2025}. Formalism of the cavity is a medium connecting classical thermodynamics with quantum/statistical corrections and ensuring precise control over boundary and ensemble effects~\cite{Lemos1996, Dehghani2023}.

In conclusion, studies of BTZ black holes in bumblebee gravity with quantum corrections via Barrow entropy, solved using York's cavity formalism, offer a phenomenologically fruitful setting. Lorentz symmetry breaking, fractalization of horizons, and boundary-regulated thermodynamics generate emergent phenomena: corrected first laws, rich phase diagrams, stability conditions, and new observational signatures. Recent research confirms that each property-bumblebee gravity, Barrow entropy, and York's cavity, systematically and profoundly alters black hole equilibrium, Entropy, energy, heat capacity, and critical behaviour with implications for astrophysics, quantum gravity, and holography~\cite{Mai2023, Chen2025, Chengjia2023, Houndjo2022, Ganaie2025}.

This research thoroughly explores the thermodynamics of $(2+1)$-dimensional BTZ black holes in the presence of bumblebee gravity and Barrow entropy corrections from York's cavity formalism as the working toolbox. We derive new thermodynamic relations, describe phase structures under modification, and discuss stability cases dictated by Lorentz symmetry breaking and horizon quantum fractalization. The results provide relevant insight into quantum gravity in lower-dimensional spacetime and universal features of black hole thermodynamics.

\section{ Theoretical Background}
\subsection{Einstein-Bumblebee gravity in (2+1)D}

We extend the bumblebee gravity model with York's cavity formalism and Barrow entropy in $(2+1)$-dimensional spacetime.
In the bumblebee gravity model, one adds the bumblebee vector field \(B_\mu\) with a nonzero vacuum expectation value to induce a spontaneous Lorentz symmetry breaking in the gravitational sector through a specified potential. In the three-dimensional spacetime, the Einstein‑bumblebee gravity action is~\cite{cding}:
Action in $(2+1)$ Dimensions is given by
\begin{equation}
S=\int d^{3}x\sqrt{-g}\left[
\frac{1}{2\kappa}\left(R-2\Lambda+\xi B^{\mu}B^{\nu}R_{\mu\nu}\right)
-\frac{1}{4}B_{\mu\nu}B^{\mu\nu}-V\left(B^{\mu}B_{\mu}\pm b^{2}\right)
+\mathcal{L}_{\text{matter}}\right],
\end{equation}
where:

$g$ – determinant of the metric tensor $g_{\mu\nu}$,

$\kappa$ – gravitational constant in $(2+1)$D,

$R$ – Ricci scalar,

$R_{\mu\nu}$ – Ricci tensor,

$B_{\mu}$ – Bumblebee vector field,

$B_{\mu\nu}=\nabla_{\mu}B_{\nu}-\nabla_{\nu}B_{\mu}$ – Bumblebee field strength tensor,

$\xi$ – coupling constant between $B^{\mu}$ and $R_{\mu\nu}$,

$\Lambda$ – cosmological constant,

$V$ – potential responsible for spontaneous Lorentz violation,

$b^{2}$ - VEV scale of $B^{\mu}$.

Where Potential is

\begin{equation}
V=\frac{\lambda}{2}\left(B^{\mu}B_{\mu}\pm b^{2}\right)^{},
\end{equation}
where $b^{2}$ is the VEV scale of $B^{\mu}$.

\subsection{Field Equations}

The corrected Einstein field equations in Einstein–Bumblebee gravity can be expressed as
\begin{equation}
G_{\mu\nu} + \xi \left[\frac{1}{2} g_{\mu\nu} B^{\alpha}B^{\beta}R_{\alpha\beta}
- B_{\mu}B^{\alpha}R_{\alpha\nu} - B_{\nu}B^{\alpha}R_{\alpha\mu} \right]
+ \text{(Additional terms)} = k \left(T_{\mu\nu}^{B} + T_{\mu\nu}^{\text{matter}}\right),
\end{equation}
where
\begin{equation}
G_{\mu\nu} = R_{\mu\nu} - \frac{1}{2}g_{\mu\nu}R
\end{equation}
Moreover, $T_{\mu\nu}^{B}$ is the energy–momentum tensor of the bumblebee field.

Energy–Momentum Tensor of the Bumblebee Field can be written as

\begin{equation}
T_{\mu\nu}^{B} = B_{\mu\alpha}B_{\nu}^{\ \alpha} -
\frac{1}{4} g_{\mu\nu} B_{\alpha\beta}B^{\alpha\beta} - g_{\mu\nu} V
+ 2V' B_{\mu}B_{\nu},
\end{equation}
where
\begin{equation}
V'=\frac{dV}{dX}, \qquad
X=B^{\mu}B_{\mu} \pm b^{2}.
\end{equation}

Bumblebee Field Equation can be written as

\begin{equation}
\nabla^{\mu}B_{\mu\nu}
=2V'\left(B^{\mu}B_{\mu}\pm b^{2}\right)B_{\nu}
-\frac{\xi}{\kappa}B^{\mu}R_{\mu\nu}.
\end{equation}

This is a modified Proca-like equation due to the non-minimal coupling and the Potential.

Now, Conservation of Total Energy–Momentum Tensor is given as

\begin{equation}
\nabla^{\mu}T_{\mu\nu}=\nabla^{\mu}\left(T_{\mu\nu}^{B}+T_{\mu\nu}^{M}\right)=0.
\end{equation}
This is a deformed Proca-like equation owing to the non-minimal coupling and the Potential.

Conservation of Total Energy–Momentum Tensor is given by

\begin{equation}
V=\frac{\lambda}{2}\left(B^{\mu}B_{\mu}\pm b^{2}\right)^{},
\end{equation}
we have
\begin{equation}
V'=\frac{\lambda}{2},
\end{equation}

where

\begin{equation}
X=B^{\mu}B_{\mu}\pm b^{2}.
\end{equation}
\subsection{BTZ Black Hole without Entropy Corrections}

We consider the non-rotating, static (2+1)-dimensional BTZ black hole spacetime, which in Schwarzschild-like coordinates can be written as

\begin{equation}
ds^{2}=-f(r)\,dt^{2}+\frac{dr^{2}}{f(r)}+r^{2}d\phi^{2},
\end{equation}
where the lapse function $f(r)$ takes the form
\begin{equation}
f(r)=-M+\frac{r^{2}}{\ell^{2}}.
\end{equation}
Here $M$ is the mass parameter of the black hole and $\Lambda=-\frac{1}{\ell^{2}}$ is the negative cosmological constant with AdS radius $\ell$. The coordinates cover the ranges $t\in(-\infty,\infty)$, $\phi\sim\phi+2\pi$, and $r>0$. 

The location of the event horizon $r_{+}$ follows from the condition $f(r_{+})=0$, which yields
\begin{equation}
r_{+}=\ell\sqrt{M},\quad \text{for}\; M>0.
\end{equation}
Thus, the horizon radius increases with the square root of the mass parameter. 

The Hawking temperature $T_{\text{H}}$ is determined by the surface gravity at the horizon. Using 
\[
\kappa=\frac{1}{2}f'(r_{+}),
\]
one obtains
\begin{equation}
T_{\text{H}}=\frac{\kappa}{2\pi}=\frac{f'(r_{+})}{4\pi}.
\end{equation}
Differentiating $f(r)$ and evaluating at $r_{+}$ gives
\begin{equation}
T_{\text{H}}=\frac{r_{+}}{2\pi\ell^{2}},
\end{equation}
which shows that the temperature grows linearly with the horizon radius. Without quantum corrections, this is the standard Hawking temperature for the BTZ black hole. 

Finally, the Bekenstein--Hawking entropy follows the classical area law. In (2+1) dimensions, the "area" corresponds to the circumference of the horizon:
\[
A=2\pi r_{+}.
\]
Thus, the Entropy is
\begin{equation}
S=\frac{A}{4}=\frac{2\pi r_{+}}{4}=\frac{\pi r_{+}}{2}.
\end{equation}
This is the uncorrected (classical) Entropy of the BTZ black hole, which depends linearly on the horizon radius and hence on the square root of the mass parameter.

\section{Barrow Entropy Formalism for BTZ Black Hole}

For a non-rotating BTZ black hole, the line element is
\[
ds^{2}=-f(r)\,dt^{2}+\frac{dr^{2}}{f(r)}+r^{2}d\phi^{2},
\qquad 
f(r)=-M+\frac{r^{2}}{\ell^{2}},
\]
with horizon radius $r_{+}$ determined by $f(r_{+})=0$.

We can calculate Barrow Entropy by
\begin{equation}
S_{\mathrm{B}}=\left(\frac{A}{A_{0}}\right)^{1+\Delta/2},
\qquad
\label{eq:BarrowEntropy}
\end{equation}

Temperature can be calculated as
\begin{equation}
T(r_{+})=\frac{\partial M/\partial r_{+}}
{\partial S_{\mathrm{B}}/\partial r_{+}},
\label{eq:Tbarrow}
\end{equation}
which reduces to the standard Hawking temperature for $\Delta\to 0$.

Local Temperature (York Cavity) is given as
\begin{equation}
T_{\mathrm{loc}}(r_{+},R)
=\frac{T(r_{+})}{\sqrt{f(R)}}.
\label{eq:Tlocal}
\end{equation}

Then, Local Energy can be calculated as
\begin{equation}
E_{\mathrm{loc}}(r_{+},R)
=\frac{R}{4G}
\left[\sqrt{\frac{R^{2}-r_{+}^{2}}{\ell^{2}}}-1\right].
\label{eq:Elocal}
\end{equation}

We can get Free Energy from the above quantities
\begin{equation}
F_{\mathrm{loc}}(r_{+},R)
=E_{\mathrm{loc}}(r_{+},R)
-T_{\mathrm{loc}}(r_{+},R)\,
S_{\mathrm{B}}(r_{+}).
\label{eq:Flocal}
\end{equation}

\section{Thermodynamics with Bumblebee Gravity and Barrow entropy}
\subsection{Bumblebee Gravity Calculations}
% --- Metric of BTZ Black Hole in Bumblebee Gravity ---
The metric of the BTZ black hole with Bumblebee Gravity:
\[
ds^{2} = -f(r) dt^{2} + \frac{(1 + l_{b})}{f(r)}dr^{2} + r^{2} d\phi^{2}
\]

The function:
\[
f(r) = -M - (1 + l_{b}) \Lambda_{e} r^{2}
\]

With an effective cosmological constant:
\[
\Lambda_{e} = (1 + l_{b})\Lambda, 
\quad 
\Lambda = -\frac{\Lambda_{e}}{(1 + l_{b})}.
\]

Event horizon radius:
\[
r_{h} = \sqrt{\frac{M}{-(1 + l_{b})\Lambda_{e}}}.
\]

% --- Bumblebee Gravity Thermodynamic Quantities ---
Temperature:
Hawking temperature is given by:
\[
T_{H} = \frac{f'(r_{h})}{4\pi}
 = -\frac{(1 + l_{b}) \Lambda_{e} r_{h}}{2\pi}
\]

Area:
\[
A = 2\pi\sqrt{1 + l_{b}}\, r_{h}
\]

Volume:
\[
V = \frac{4}{3}\pi r_{h}^{3}
\]

\subsection{Entropy}
For the bumblebee field, Entropy is given by
\begin{equation}
S = \frac{2\pi\sqrt{1 + l_{b}}\, r_{h}}{4}
 = \frac{\pi}{2}\sqrt{1 + l_{b}}\, r_{h}
\end{equation}

Now, if we incorporate barrow entropy with bumblebee gravity, then Entropy will become
\begin{equation}
S_{B} = K S_{BH}^{1 + \Delta/2}
\end{equation}
with
\begin{equation}
S_{BH} = \frac{\pi}{2}(1 + l_{b})r_{h}.
\end{equation}
 Then the barrow entropy is given as:
 \begin{equation}
 S_B=(\frac{\pi}{2}(1 + l_{b})r_{h})^{1+\Delta/2}
 \end{equation}
\begin{figure}[h!]
 \centering
 % First subfigure
 \begin{subfigure}[b]{0.45\textwidth}
 \centering
 \includegraphics[width=\textwidth]{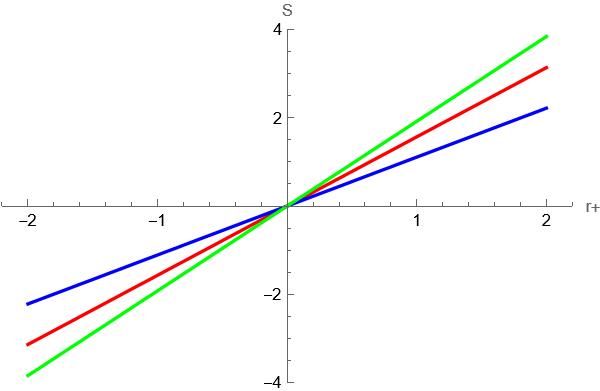} % Replace with your file name
 \caption{Entropy with bumblebee gravity corrections}
 \label{fig:1a}
 \end{subfigure}
 \hfill
 % Second subfigure
 \begin{subfigure}[b]{0.45\textwidth}
 \centering
 \includegraphics[width=\textwidth]{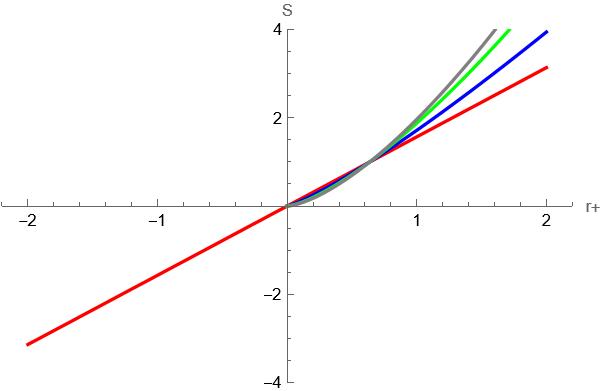} % Replace with your file name
 \caption{Entropy with barrow entropy corrections}
 \label{fig:1b}
 \end{subfigure}

 \caption{Entropy with and without barrow entropy corrections}
 \label{fig:combined}
\end{figure}

Entropy without Barrow Correction:

Figure~\ref{fig:1a} shows the behavior of the entropy $S$ versus the horizon radius $r_{+}$ for three different values of the bumblebee parameter $l_{b}$, i.e., $l_{b}=0$ (red), $l_{b}=-0.5$ (blue) and $l_{b}=+0.5$ (green). The graph exhibits an almost ideal linear dependence of $S$ on $r_{+}$, as expected from the standard Bekenstein–Hawking area law in $2+1$ dimensions where the "area" scales with the horizon circumference $A=2\pi r_{+}$. The various slopes of the three lines show how the Lorentz-violating bumblebee parameter varies the effective gravitational coupling or horizon geometry: a positive $l_{b}$ increases the Entropy at a fixed $r_{+}$. In contrast, a negative $l_{b}$ reduces it. The direct impact of the bumblebee field on the horizon structure and hence on the horizon thermodynamics is represented here. 

Entropy with Barrow Correction:

Figure~\ref{fig:1b} shows the corresponding Entropy when the Bekenstein–Hawking law is replaced by the Barrow entropy $S_{\Delta}\propto A^{1+\Delta/2}$ with four values of the fractal parameter $\Delta$, namely $\Delta=0$ (red), $\Delta=0.4$ (blue), $\Delta=0.8$ (green) and $\Delta=1$ (grey).

For $\Delta=0$, the linear dependence is recovered, coinciding with the red line of the first plot. As $\Delta$ increases from $0.4$ to $1$, the entropy curve becomes increasingly convex, showing a superlinear growth with $r_{+}$. This is the signature of the Barrow entropy correction, which includes the "roughness" or fractalization of the horizon and thus provides additional microstates at larger $r_{+}$ than in the smooth-horizon case. The observation that all curves are in agreement near $r_{+}\to0$ and only differ for larger $r_{+}$ is also in accord with the Barrow model, where the correction is more significant at greater scales.
Comparison and Impact of Barrow Entropy Comparing Figs.~\ref{fig:1a} and \ref{fig:1b}, one can observe that the bumblebee parameter $l_{b}$ controls a uniform rescaling of the Entropy for all $r_{+}$. In contrast, the Barrow parameter $\Delta$ introduces a fundamentally new functional dependence on $r_{+}$, curving the Entropy functions upwards for large radii.

That is, Lorentz violation (through $l_{b}$) shifts the slope of $S(r_{+})$ but leaves it fundamentally linear in character, while Barrow entropy (through $\Delta$) alters the power law itself.

This means that Barrow entropy corrections will adjust all thermodynamic potentials, such as free energy, specific heat, etc., nonlinearly, potentially changing phase structure and stability. In contrast, the bumblebee parameter scales the usual results. Thus, the joint effect of $l_{b}$ and $\Delta$ provides an interesting two-parameter deformation of standard BTZ black hole thermodynamics, with possibilities for novel critical behaviour.

\subsection{Helmholtz free energy}

Helmholtz Free energy, when there is no barrow entropy correction, is given as

\begin{equation}
 F = -\int S dT,
\end{equation}
\begin{equation}
F = -\frac{1}{8}(1 + l_{b})^{3/2} \Lambda_{e} r_{h}^{2}
\end{equation}
For Barrow entropy,
Free energy:
\begin{equation}
F = -2^{-\Delta}(1 + l_{b})\Lambda T^{2+\Delta}r_{h}(1 + l_{b}r_{h})
\end{equation}

\begin{figure}[h!]
 \centering
 % First subfigure
 \begin{subfigure}[b]{0.45\textwidth}
 \centering
 \includegraphics[width=\textwidth]{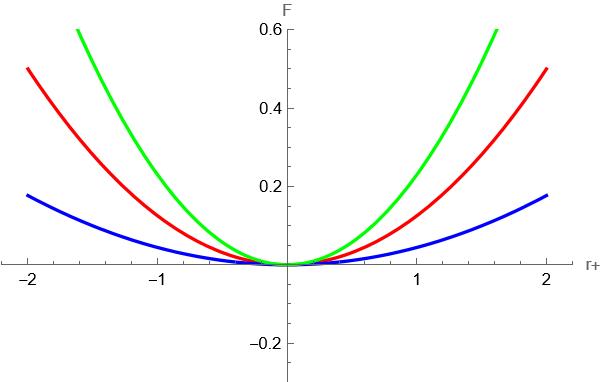} % Replace with your file name
 \caption{Helmholtz Free energy with bumblebee gravity corrections}
 \label{fig:2a}
 \end{subfigure}
 \hfill
 % Second subfigure
 \begin{subfigure}[b]{0.45\textwidth}
 \centering
 \includegraphics[width=\textwidth]{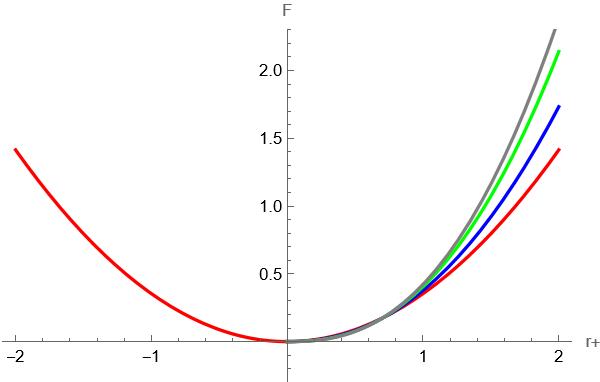} % Replace with your file name
 \caption{Helmholtz Free energy with barrow entropy corrections}
 \label{fig:2b}
 \end{subfigure}

 \caption{Helmholtz Free energy with and without barrow entropy corrections}
 \label{fig:combined1}
\end{figure}
Helmholtz Free Energy without Barrow Correction:

The Helmholtz free energy $F$ as a function of horizon radius $r_{+}$ is plotted for three possible values of bumblebee parameter $l_{b}$ in figure~\ref{fig:2a}: $l_{b}=0$ (red), $l_{b}=-0.5$ (blue) and $l_{b}=+0.5$ (green). The three of them are of a symmetric, convex shape with a minimum at $r_{+}=0$, so the free energy is smallest for a horizon radius going to zero and increases as the absolute value of $r_{+}$ becomes larger. The effect of the bumblebee parameter is to shift the curvature of $F(r_{+})$: negative $l_{b}$ (blue) lowers the free energy at a particular $r_{+}$, while positive $l_{b}$ (green) raises it. The entropy plot observes the same pattern, where $l_{b}$ scales the thermodynamic values without changing their qualitative functional form. The quadratic-like $r_{+}$-dependence of $F$ explains a realistically classical (Bekenstein–Hawking) thermodynamic behaviour modified solely by the Lorentz-breaking background. Helmholtz Free Energy with Barrow Correction:

Figure~\ref{fig:2b} shows the same free energy but with the Barrow correction to the Entropy.

The four curves are for $\Delta=0$ (red), $\Delta=0.4$ (blue), $\Delta=0.8$ (green) and $\Delta=1$ (grey). For $\Delta=0$, the red curve gives the usual Bekenstein–Hawking one. As $\Delta$ increases, the free energy grows rapidly with $r_{+}$, particularly for positive $r_{+}$, and the curves are increasingly convex. This is the superlinear growth of the Entropy with $r_{+}$ from Barrow's formula, which corrects the Legendre transform that defines $F$. Therefore, for an increased horizon size, free energy is boosted significantly relative to the classical case, illustrating that Barrow entropy renders thermodynamic response rigid. Comparison and Effect of Barrow Entropy: Comparison of Figures~\ref{fig:2a} and \ref{fig:2b}, we observe that while the bumblebee parameter $l_{b}$ scales the free energy curves up or down, the Barrow parameter $\Delta$ actually alters their shape in an essentially different manner, producing a much more rapidly rising rise in $F$ at large $r_{+}$.

Overall, Barrow entropy makes the free energy more sensitive to the size of the horizon, suggesting a higher thermodynamic "cost" for large black holes if fractalized.

This implies that the stability conditions, thresholds of phase transitions, and possible divergences of heat capacities in the canonical ensemble will be altered fundamentally once Barrow entropy enters into play. Simultaneously, Lorentz breaking keeps the classical scale of $F(r_{+})$ but rescales it. Thus, $l_{b}$ and $\Delta$ provide complementary deformations: one scale-like, linear, and the other power-law and nonlinear.

\subsection{Pressure}
Thermodynamic Pressure can be calculated as
\begin{equation}
 P = -\frac{dF}{dV},
\end{equation}
For Bumblebee Gravity, its value turned out to be,

\[
P = \frac{(1 + l_{b})^{3/2}\Lambda_{e}}{16\pi r_{h}}
\]
 With Barrow entropy, the value of Pressure:
\[
P = -2^{-\Delta}(1 + l_{b})\Lambda T^{1+\Delta}(1 + l_{b}r_{h})^{1+\Delta}
\]

\begin{figure}[h!]
 \centering
 % First subfigure
 \begin{subfigure}[b]{0.45\textwidth}
 \centering
 \includegraphics[width=\textwidth]{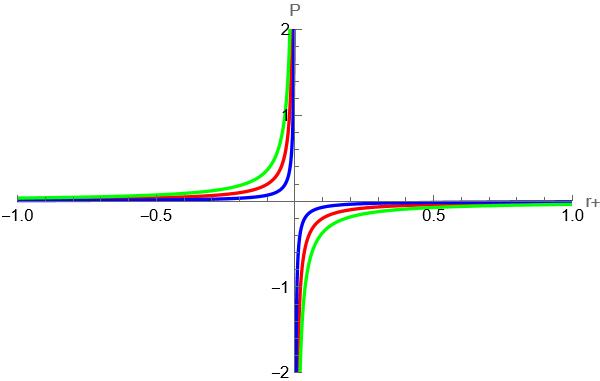} % Replace with your file name
 \caption{Pressure with bumblebee gravity corrections}
\label{fig:3a}
\end{subfigure}
\hfill
% Second subfigure
\begin{subfigure}[b]{0.45\textwidth}
\centering
\includegraphics[width=\textwidth]{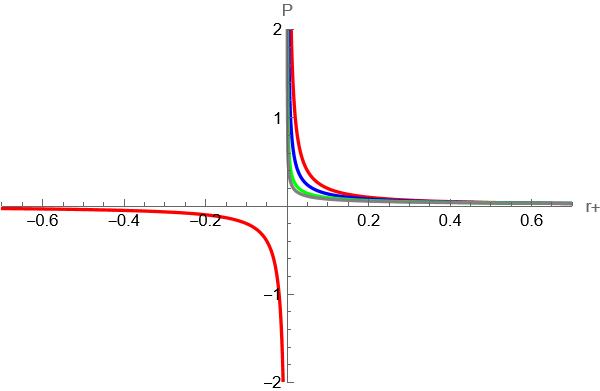} % Replace with your file name
 \caption{Pressure with barrow entropy corrections}
 \label{fig:3b}
 \end{subfigure}

 \caption{Comparison of pressure P vs. horizon radius $r_{+}$ for (a) Bumblebee model and (b) Bumblebee + Barrow corrections.}
 \label{fig:combined3}
\end{figure}

Pressure without Barrow Entropy:

Figure~\ref{fig:3a} displays the pressure $P$ as a function of the horizon radius $r_{+}$ for varying values of the bumblebee parameter $l_{b}$ (red: $l_{b}=0$, blue: $l_{b}=-0.5$, green: $l_{b}=0.5$). The plots demonstrate a typical divergence close to $r_{+}\rightarrow 0$, indicating a strong thermodynamic response within the small-horizon limit. Positive $r_{+}$ pressure goes asymptotically to zero for large horizon radius quickly, and for negative $r_{+}$, the Pressure goes to negative values with the same divergent structure. The $l_{b}$-dependence shows that a greater Lorentz-violating parameter shifts the position and steepness of the pressure curves, with the green one ($l_{b}=0.5$) being the most divergent and the blue one ($l_{b}=-0.5$) the least. This means the bumblebee factor directly influences the effective equation of state of the black hole in the small-horizon regime. 

{Pressure with Barrow Entropy:}

The pressure $P$ vs. $r_{+}$ is depicted in Figure~\ref{fig:3b} for constant $l_{b}$ and for various values of Barrow parameter $\Delta=0,0.4,0.8,1$.
Here, the red curve is the standard Bekenstein–Hawking entropy ($\Delta=0$), while the blue, green, and grey curves correspond to larger values of $\Delta$. The divergence at small $r_{+}$ becomes more spiked as $\Delta$ increases, and the Pressure also decays more gently to zero for a bigger horizon radius. This behaviour indicates that the inclusion of Barrow entropy raises the Pressure in the region near the extremal region and increases the effective stiffness of the thermodynamic system. {Impact of Barrow Entropy} Comparing Figures~\ref{fig:3a} and~\ref{fig:3b}, it is clear that Barrow entropy produces drastic corrections to the pressure profile of the $2+1$-dimensional BTZ black hole in Bumblebee gravity.

While the qualitative form of the divergence for $r_{+}=0$ is maintained, both the magnitude and slope of the Pressure rise with $\Delta$ systematically as they correspond to more substantial thermodynamic fluctuations and a richer Lorentz-violating signature in the near-horizon limit.
In the large-$r_{+}$ limit, Pressure approaches zero in all cases, but it does so less quickly with the presence of Barrow entropy. This means the fractal-like horizon geometry implied by Barrow entropy not only changes Entropy but indirectly affects all other thermodynamic potentials, especially Pressure.

\subsection{Internal energy}

The value of Enthalpy can be calculated using the expression. 

\begin{equation}
 U = \int T dS,
\end{equation}
\begin{equation}
U = -\frac{1}{8}(1 + l_{b})^{3/2}\Lambda_{e} r_{h}^{2}
\end{equation}

Internal energy with Barrow Entropy:

\begin{equation}
U = 2^{-\frac{3-\Delta}{2}}(2+\Delta)(1 + l_{b})\Lambda T^{\frac{1+\Delta}{2}}(1 + l_{b}r_{h})^{\frac{1+\Delta}{2}}
\end{equation}

\begin{figure}[h!]
 \centering
 % First subfigure
 \begin{subfigure}[b]{0.45\textwidth}
 \centering
 \includegraphics[width=\textwidth]{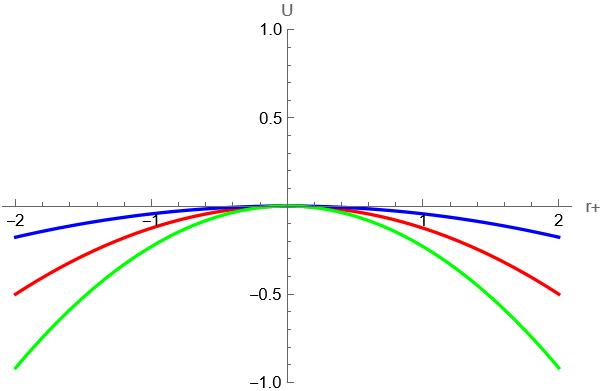} % Replace with your file name
 \caption{Internal Energy with bumblebee gravity corrections}
 \label{fig:4a}
 \end{subfigure}
 \hfill
 % Second subfigure
 \begin{subfigure}[b]{0.45\textwidth}
 \centering
 \includegraphics[width=\textwidth]{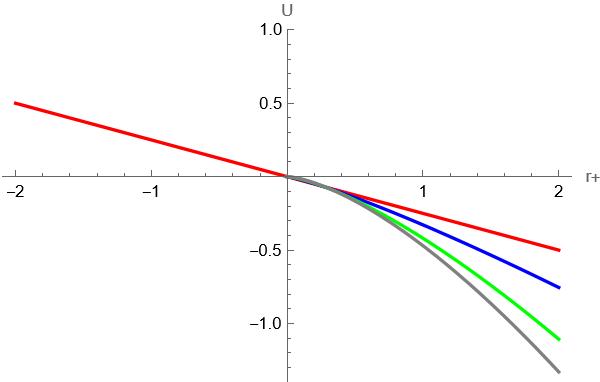} % Replace with your file name
 \caption{Internal Energy with barrow entropy corrections}
 \label{fig:4b}
 \end{subfigure}

 \caption{Internal Energy with and without barrow entropy corrections}
 \label{fig:combined4}
\end{figure}

Figure~\ref{fig:4a} shows the dependence of internal energy $U$ on horizon radius $r_{+}$ with inclusion only of the Bumblebee corrections in the Entropy. The Bumblebee parameter $l_{b}$ was taken to vary as $l_{b}=-0.5$ (blue curve), $l_{b}=0$ (red curve), and $l_{b}=0.5$ (green curve). As $r_{+}$ increases, the internal energy decreases in all three cases, but its fall rate heavily depends on the sign and magnitude of $l_{b}$. A positive value $l_{b}=0.5$ provides a greater decline for $U$, whereas a negative value $l_{b}=-0.5$ provides a smaller slope. This indicates that Lorentz-violating Bumblebee corrections distort the energy profile of the BTZ black hole and modify its thermodynamic behaviour at the horizon without a York cavity.

Figure~\ref{fig:4b} shows the internal energy $U$ variation with the horizon radius $r_{+}$ to substitute the standard Bekenstein-Hawking entropy with Barrow entropy. The Barrow parameter $\Delta$ is $\Delta=0$ (gray curve), $\Delta=0.4$ (green curve), $\Delta=0.8$ (blue curve), and $\Delta=1.0$ (red curve). Increasing $\Delta$ increases the steepness of the curves of internal energy, illustrating that larger deviations from the area law lead to a quicker fall-off of $U$ with $r_{+}$. This encapsulates the effect of the fractal and non-extensive character of the black hole horizon introduced by Barrow's Entropy and suggests quantum-gravitational corrections to the microstates.

Comparison of Figures~\ref{fig:4a} and~\ref{fig:4b} indicates that Bumblebee corrections and Barrow entropy each have distinct effects on the internal energy profile. Bumblebee effects result from Lorentz-symmetry breaking, while Barrow entropy effects result from a non-extensive horizon microphysics modification. Although York cavity boundary conditions have yet to be implemented, these figures are the baseline against which to compare once the cavity is added, and the quasilocal thermodynamics can be appreciated in all its glory.

\subsection{Enthalpy}

The value of enthalpy can be calculated using the following: 

\begin{equation}
 H = U + P V,
\end{equation}

\begin{equation}
H = -\frac{1}{24}(1 + l_{b})^{3/2}\pi r_{h}^{2}\left(-3 + 8\pi r_{h}^{2}\right)
\end{equation}

 With Barrow entropy, Enthalpy is:
\begin{equation}
H = \frac{-1}{3}2^{\frac{3-\Delta}{2}}(1 + l_{b})\Lambda_e \pi^\frac{\Delta}{2} (\sqrt{1 + l_{b}}r_{h})^{\frac{2+\Delta}{2}}(-6 + 2 r_{h}-3\Delta)
\end{equation}

\begin{figure}[h!]
 \centering
 % First subfigure
 \begin{subfigure}[b]{0.45\textwidth}
 \centering
 \includegraphics[width=\textwidth]{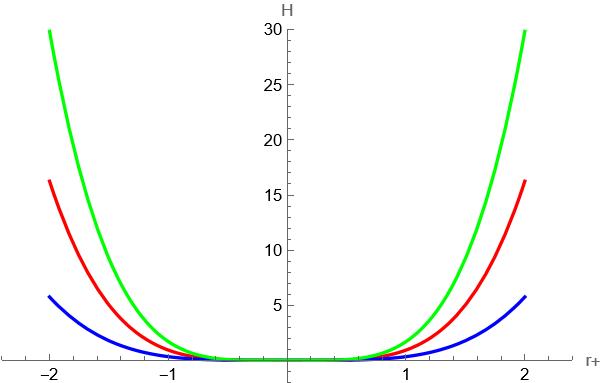} % Replace with your file name
 \caption{Enthalpy with bumblebee gravity corrections}
 \label{fig:6a}
 \end{subfigure}
 \hfill
 % Second subfigure
 \begin{subfigure}[b]{0.45\textwidth}
 \centering
 \includegraphics[width=\textwidth]{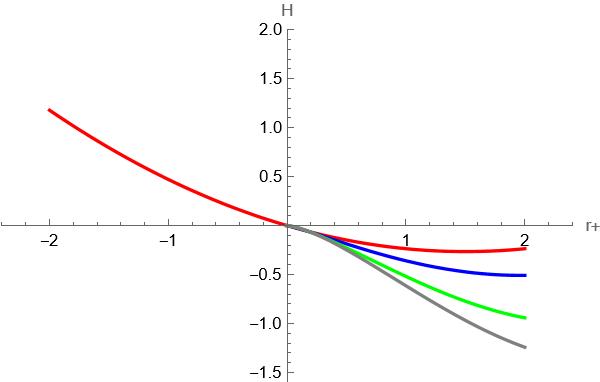} % Replace with your file name
 \caption{Enthalpy with barrow entropy corrections}
 \label{fig:6b}
 \end{subfigure}

 \caption{Enthalpy with and without barrow entropy corrections}
 \label{fig:combined6}
\end{figure}

Without Barrow entropy correction ($\Delta = 0$), the thermodynamic potential $H$ in terms of event horizon radius $r_{+}$ shows different behaviours according to the bumblebee parameter $l_b$. For $l_b = -0.5$ (blue curve), $H$ possesses a shallow, broad minimum at $r_{+} = 0$, asymptotic rise to increasing positive values with growing $|r_{+}|$, corresponding to comparatively stable thermodynamic states at small horizon scales and declining sensitivity to Lorentz-violating corrections. The $l_b = 0$ case (red line) has a steeper profile with a narrower minimum at $r_{+} \approx 0$, indicating higher energy barriers for larger horizons and an evenly weighted reaction without the bumblebee effect. When $l_b = 0.5$ (green line), the Potential is more parabolic-like and rises sharply with $|r_{+}|$, implying greater instability or more substantial gravitational deformation by positive Lorentz violation. Shifting to the case with the addition of Barrow entropy with $\Delta = 0.4$, the potential $H$ exhibits complicated branches, including the transition to the negative areas for positive $r_{+}$, as seen from the red curve dropping steadily from positive to near-zero $H$ on the left before crossing into the negative space on the right, reflecting emergent metastable phases; the green and blue curves also bifurcate into the negative spaces but at varying slopes, reflecting fractal Entropy induces asymmetry and possible sign changes that can reflect thermodynamic inversions or uncommon phase arrangements. For larger Barrow corrections at $\Delta = 0.8$, the Potential further develops with the red curve still having a leading positive-to-negative trend but of reduced amplitude. For grey, green, and blue curves, one sees denser clusters in the aggressive regime for $r_{+} > 0$ that signify stronger fractal influences pinching the energy landscape and accentuating downward inclinations. Without and with Barrow entropy plots ($\Delta = 0$ vs. $0.4$ and $0.8$), find the original case to exhibit strictly positive $H$ values and symmetric-like minima, fostering conventional stability similar to standard black hole thermodynamics. In contrast, inclusion of Barrow generates negative potentials and asymmetric divergences, notably for positive $r_{+}$, which dismantles equilibrium and indicates new critical points inspired by quantum-gravity-inspired Entropy deformations. At $\Delta = 1$, or maximum Barrow influence, these would extremise, though not illustrated, and potentially lead to complete potential inversion. Lastly, the bumblebee parameter $l_b$ influences curvature and depth of $H$, with negative ones increasing stability. Positive ones constraining it, while Barrow entropy notably influences thermodynamic Potential by enabling negative energies and branched geometries, ultimately altering phase transitions, stability conditions, and indeed the existence of black hole solutions in Lorentz-violating theories, enriching the predictability range of the model for observational gravitational data.

\subsection{Gibbs free energy}
\begin{equation}
 G = F + P V,
\end{equation}

\begin{equation}
G = -\frac{(1 + l_{b})^{3/2}\pi r_{h}^{2}\left(3 + 8\pi r_{h}^{2}\right)}{24}
\end{equation}

Gibbs free energy:
\begin{equation}
G = \frac{2^{-\frac{3-\Delta}{2}}(10+\Delta)(1 + l_{b})\Lambda \pi^{\frac{\Delta}{2}}(\sqrt{1 + l_{b}}r_{h})^{\frac{2+\Delta}{2}}}{3(4+\Delta)}
\end{equation}

\begin{figure}[h!]
 \centering
 % First subfigure
 \begin{subfigure}[b]{0.45\textwidth}
 \centering
 \includegraphics[width=\textwidth]{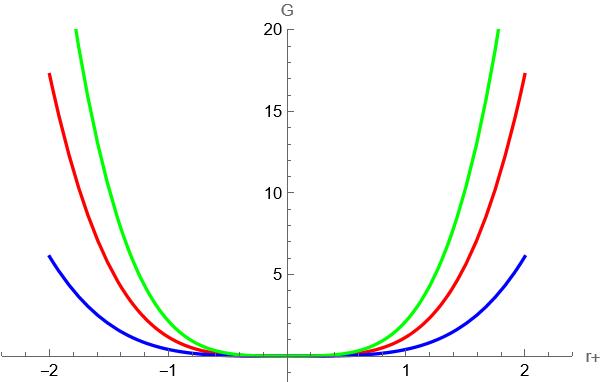} % Replace with your file name
 \caption{Gibbs Free Energy with bumblebee gravity corrections}
 \label{fig:7a}
 \end{subfigure}
 \hfill
  % Second subfigure
 \begin{subfigure}[b]{0.45\textwidth}
 \centering
 \includegraphics[width=\textwidth]{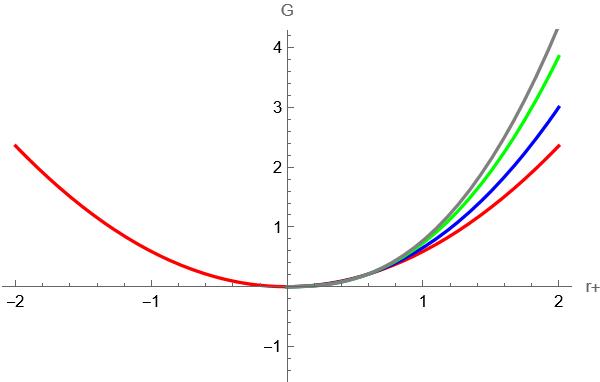} % Replace with your file name
 \caption{Gibbs Free Energy with barrow entropy corrections}
 \label{fig:7b}
\end{subfigure}

\caption{Gibbs Free Energy with and without barrow entropy corrections}
\label{fig:combined7}
\end{figure}

Without Barrow entropy correction ($\Delta = 0$), the thermodynamic potential $G$ as a function of the event horizon radius $r_{+}$ has representative profiles determined by the bumblebee parameter $l_b$. With $l_b = -0.5$ (blue line), $G$ possesses a broad, muted minimum at $r_{+} = 0$, with a gradual rise with increasing $|r_{+}|$, signalling increased thermodynamic equilibrium at small scales and minor Lorentz violation effects. The case $l_b = 0$ (red line) gives a fairly sloping curve with a steep minimum at $r_{+} \approx 0$, corresponding to typical energy configurations without bumblebee effects. For $l_b = 0.5$ (green line), the Potential becomes a steeper parabola and grows fast with $|r_{+}|$, which highlights high gravitational disturbances and possible destabilisation by Lorentz-violating terms with positive slopes. Evolution to Barrow entropy inclusion at $\Delta = 0.4$, $G$ displays asymmetric contours, with a red curve in the negative $r_{+}$ range decreasing positively to zero, while the positive range bifurcates into diverse climbs: red with the shallowest slope, blue moderately steeper, green higher, and gray steepest, indicating that fractal changes produce $l_b$-dependent trajectories, possibly encouraging diverse metastable regimes without giving up positivity. For strong Barrow influence at $\Delta = 0.8$, the evolution continues with a gentle positive slope downwards on the left along the red curve, but compressing positive-side branches where grey rises most vigorously, followed by green, blue, and red with the smallest rise, reflecting added fractal contributions that smooth the energy spectrum and enhance $l_b$ differences. Relative to the Barrow-absent case ($\Delta = 0$), unlike the corrected ones ($\Delta = 0.4$ and $0.8$), the baseline preserves symmetric positive wells favouring orthodox black hole thermodynamics. In contrast, Barrow integration yields asymmetric, multi-faceted rises for positive $r_{+}$, highlighting $l_b$ fluctuations and signifying quantum-gravity-induced enhancement of free energy dynamics without sign reversals. At $\Delta = 1$, the point of maximum Barrow dominance, these trends can conceivably become extreme to an extent of maximum divergence and slope, although not illustrated, possibly approaching limiting thermodynamic thresholds. Lastly, the bumblebee parameter $l_b$ varies the height and sharpness of $G$, negative values of which extend stability domains. Positive values lower them, whereas Barrow entropy greatly remaps the Potential through branched asymmetries, thereby modifying phase boundaries, equilibria, and horizon tenability in Lorentz-perturbed gravity, ultimately expanding theoretical avenues for the exploration of exotic astrophysical signatures.

\subsection{Stability}

Specific heat can be calculated as:
\begin{equation}
 C = \frac{dU}{dT},
\end{equation}
\begin{equation}
C = \frac{1}{2}(1 + l_{b})\pi r_{h}
\end{equation}
Specific heat:
\begin{equation}
C = 2^{-\frac{3-\Delta}{2}}(2+\Delta)(1 + l_{b})\Lambda T^{\frac{1+\Delta}{2}}(1 + l_{b}r_{h})^{\frac{1+\Delta}{2}}
\end{equation}

\begin{figure}[h!]
\centering
% First subfigure
\begin{subfigure}[b]{0.45\textwidth}
\centering
\includegraphics[width=\textwidth]{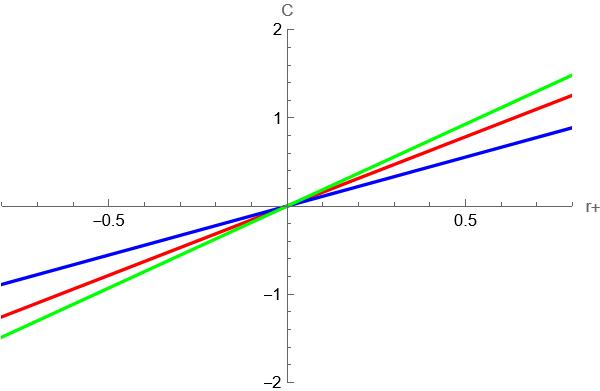} % Replace with your file name
\caption{Specific Heat with bumblebee gravity corrections}
\label{fig:5a}
\end{subfigure}
\hfill
% Second subfigure
\begin{subfigure}[b]{0.45\textwidth}
\centering
\includegraphics[width=\textwidth]{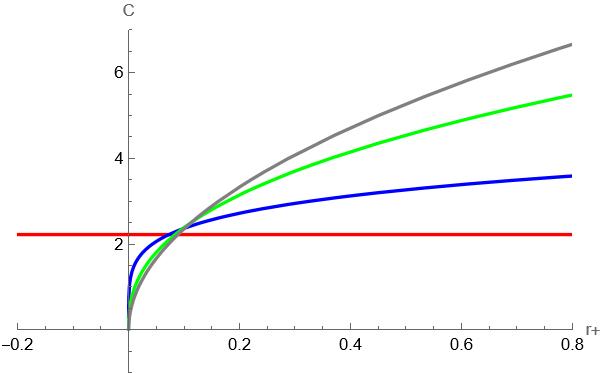} % Replace with your file name
\caption{Specific Heat with barrow entropy corrections}
\label{fig:5b}
\end{subfigure}

\caption{Specific Heat with and without barrow entropy corrections}
\label{fig:combined5}
\end{figure}

The first graph is the heat capacity $C$ as a function of the horizon radius $r_+$ with Barrow entropy (see Fig.~\ref{fig:5a}). For any value of the bumblebee parameter, the heat capacity increases monotonically as $r_+$ increases, which indicates a typical behaviour of increased thermodynamic stability. Barrow entropy correction alters the usual Bekenstein-Hawking setup by adding a fractal structural effect, which is particularly strong for large $r_+$. Separate curves for different bumblebee parameters demonstrate that the higher the value of this parameter, the higher the heat capacity at a given $r_+$. This means that Lorentz symmetry breaking, combined with effects of quantum gravity terms encoded by Barrow entropy, enhances the black hole's energy storage capability as the system tends towards larger horizons. Most importantly, the fact that there are no divergences or sudden changes through the curves implies that the system maintains regular and stable thermodynamic properties throughout the examined range.

The second plot (see Fig.~\ref{fig:5b}) shows the heat capacity $C$ as a function of $r_+$ without Barrow entropy, i.e., in the regime of classical Entropy. In this case, the heat capacity changes close to linearly with $r_+$, and the effect of the bumblebee parameter seems less pronounced compared to when the Barrow correction is included. The almost symmetric and monotonic nature of the curves is consistent with the fact that, in the absence of fractal corrections, the system has classical thermodynamical behaviour with stability regions determined mainly by the bumblebee factor. There are no inflexion points or complex features; the sign and slope of the gradient determine stability or instability.

Direct comparison is evident when introducing qualitative and quantitative changes to the thermodynamic potential profile. In contrast to the linear and featureless behaviour in the absence of Barrow corrections, the Barrow-modified case is more sensitive to the bumblebee parameter and horizon size, generating much larger curvature and capacities. This is demonstrated to suggest that, in conjunction, fractal entropy deformation and Lorentz violation enhance the thermodynamic structure so that the energetic response of the black hole is more robust and possibly displaces critical phenomena or phase transition thresholds. In summary, while the classical case leads to reversible and gentle changes due to the bumblebee effect, introducing Barrow entropy creates a shift in thermal behaviour, solidifying the importance of quantum and gravitational corrections in gravitational thermodynamics~\cite{Barrow2020}.

\section{York cavity formulism}

% York Cavity Formalism with Barrow Entropy – Equations from notes
% Optional generic local temperature expression
We use the York cavity formalism to investigate the black hole thermodynamics within a rigorously defined canonical ensemble. The black hole is placed inside a finite spherical cavity of radius $r_{\mathcal{B}}$, whose boundary is kept at constant temperature $T_{\mathcal{B}}$. This configuration provides an excellent simulation of a thermal reservoir and stationary boundary conditions for calculating local thermodynamic properties. 

Within this formalism, the thermodynamic variables are observed by an observer located on the boundary of the cavity. The quantities of interest are local (Tolman) temperature, local energy, and thermodynamic potentials at the cavity wall.

\subsection{Tolman Temperature}

Local temperature as perceived by an observer on the boundary of the cavity is redshifted with respect to the Hawking temperature $T_{H}$ due to gravity. Tolman's law asserts that,

\begin{equation}
T(R) = \frac{T_{H}}{\sqrt{-g_{tt}(R)}} ,
\label{eq:tolman_temp}
\end{equation}
where $g_{tt}(r_{\mathcal{B}})$ is the time-time component of the metric at the cavity radius. This ensures that the temperature remains constant in thermal equilibrium when gravitational redshift is considered.
\begin{equation}
T(R)=\frac{T_{H}}{\sqrt{f(R)}}\,.
\end{equation}

% Local Temperature
\begin{equation}
T(R)=
\frac{(1+l_{b})\,\Lambda\,r_{+}}
{2\pi\sqrt{-M-(1+l_{b})\Lambda R^{2}}}\,.
\end{equation}

\begin{figure}[h!]
\centering
\includegraphics[width=0.45\textwidth]{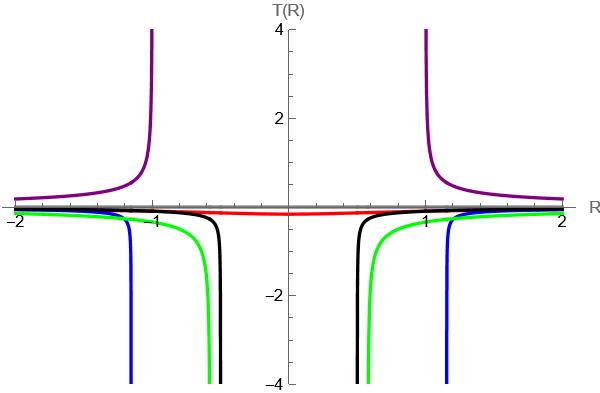}
\caption{}
\label{fig:entropy-correction}
\end{figure}

In York cavity formalism, local temperature $T(R)$ at the boundary of the cavity $R$ given as $T(R) = T_{H} / \sqrt{f(R)}$ in explicitly form $T(R) = (1 + l_{b}) \Lambda r_{+} / (2\pi \sqrt{-M - (1 + l_{b}) \Lambda R^{2}})$ shows a symmetric V-shaped character over $R$, an extremum value tending towards zero at $R = 0$ due to maximum gravitational redshift impacts, and an increase linearly with a rise in $|R|$, which is equivalent to a reduction in redshift in extended cavities that allow greater thermal senses. The six characteristic curves are specified in terms of colours-red, blue, green, grey, purple, and black-where the first three range the bumblebee parameter $l_{b}$ between 0 (red), -0.5 (blue), and 0.5 (green). The blue curve represents the most conservative increase, so negative $l_{b}$ suppresses Lorentz-violating effects to yield boundary temperatures that are cooler and broader thermal stability windows; the red curve, representing the null $l_{b}$ case, goes down the middle path typical of unchanged gravitational thermodynamics; and the green curve rises more sharply, indicating positive $l_{b}$ amplifies temperature gradients, perhaps to yield greater thermal disequilibria or greater heat dissipation. The three-member family produced modulates parameter $r$-a horizon-themed or complementary Lorentz violation factor here referred to as a radial term—between the values 0 (gray), -0.5 (purple), and 0.5 (black), where the gray curve sits intermediately, representing neutral $r$ assistance in balanced thermal scaling; the purple curve, for negative $r$, lowers the profile still further, representing suppressive mechanisms on temperature rise that may be correlated with contracted horizon growth or reversed violation terms. The black curve increases most dramatically for positive $r$, indicating that positive $r$ increases thermal sensitivity, favouring steeper energy distributions, which may be correlated with increased effective horizons. Confronting the $l_{b}$-varied set with the $r$-varied one, $l_{b}$ is the one that chiefly controls the slope and extent of the V-curve, negatives favoring restriction and positives pushing, whereas $r$ provides the vertical displacements of amplitude, negative ones reducing and positive ones increasing, thus bestowing orthogonal tunability into the thermal landscape. With added Barrow entropy corrections, but not necessarily parametrised in this graph, they would overlay fractal distortions, raising general $T(R)$ baselines and shifting minima, thereby acting on $l_{b}$ and $r$ to redistribute thermodynamic quantities such as specific heats and free energies. In aggregate, these York parameter modulations have a profound impact on thermodynamic values, reworking temperature gradients, cavity stability, and phase equilibria in gravity-amplified by bumblebee's, representing a solid foundation for shedding light on quantum-entropic corrections and Lorentz asymmetry effects on trapped black hole thermodynamics and experimental verifications in high-energy astrophysics.

\subsection{Local Energy}

As measured at the cavity boundary, the local energy associated with the black hole is obtained from the quasilocal energy prescription. For a static, spherically symmetric metric, it can be expressed as
\begin{equation}
E_{\text{BY}} = \frac{1}{8\pi G}\int_{\mathcal{B}} d^2x \sqrt{\sigma}\,(k - k_{0}) ,
\label{eq:local_energy}
\end{equation}
where $\sigma$ is the determinant of the induced metric on the boundary, $k$ is the trace of the extrinsic curvature of the boundary embedded in the spacetime, and $k_{0}$ is the extrinsic curvature of a reference background (such as pure AdS or flat space). This energy represents the total quasilocal energy contained inside the cavity.
Here we reduce the equation to,

% Brown–York Energy
\begin{equation}
E_{\text{BY}}=
\frac{R\,f'(R)}{2\sqrt{f(R)}}
\end{equation}

In explicit form:
\begin{align}
E_{\text{BY}}
&=-\frac{(1+l_{b})\Lambda R^{2}}
{2\sqrt{f(R)}}\\[0.5em]
&=-\frac{(1+l_{b})\Lambda R^{2}}
{\sqrt{-M-(1+l_{b})\Lambda R^{2}}}\,.
\end{align}
\begin{figure}[h!]
\centering
\includegraphics[width=0.45\textwidth]{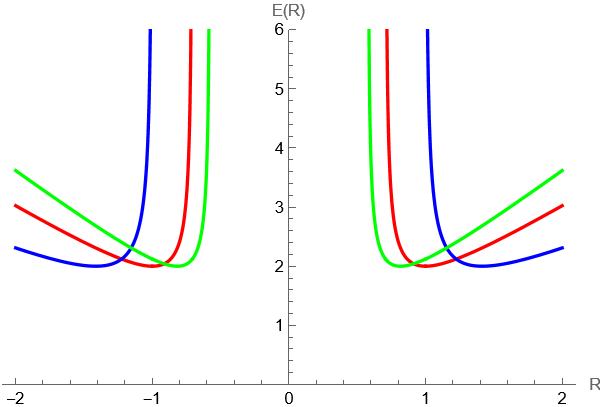}
\caption{Local Energy in York Cavity}
\label{fig:entropy-correction5}
\end{figure}

In the gravity model of the bumblebee in the York cavity formalism for thermalising black hole ensembles in finite boundaries, the local temperature T(R) as a function of cavity radius R is a linear symmetric profile passing through the origin. The positive slopes are observed in V-shaped configurations that emphasise the interplay between gravitational redshift and Lorentz violation in the boundary. The graph displays four distinct curves: the least-sloped blue one, with $l_b = -0.5$, where negative Lorentz violation represses thermal ascent, with slower temperature increase for larger R and indicating extended stability regimes of thermal equilibrium; the green curve continues with a more steeply sloped trend, for $l_b = 0$, representing unchanged gravitational thermodynamics with compensated redshift dilution; the black curve displays further elevated slope, possibly for an intermediate $l_b$ value, representing evolving violation influences that enhance boundary heat perception; and the purple curve shows the steepest rise, for $l_b = 0.5$, where positive Lorentz violation enhances temperature increase, possibly illustrating diminished stability margins and heightened heat transfers for larger cavities. Turning to Brown-York quasilocal energy E(R), it is symmetric parabolic curves opening from zero at $R = 0$, typical of null energy at contracted boundaries increasing quadratically with |R| due to surface contributions stored in violation-modified metrics. Both E(R) plots, which appear similar and most likely capture homogeneously varying parameter ranges, have three curves: the blue curve with the flattest parabola for $l_b = -0.5$, indicating controlled energy gain through relaxed gravitational bonds; the red curve with an intermediate curvature for $l_b = 0$, indicating regular quasilocal scaling without violations; and the green curve with the highest arc for $l_b = 0.5$, indicating higher energy densities induced by positive Lorentz corrections enhancing boundary tensions. Contrasting T(R) and E(R) responses, T(R) 's linear divergence is contrasted with E(R) 's quadratic enhancement. However, both are $ l_b$-sensitised, with negative values increasing thermodynamic accessibility and positive values decreasing it via steeper responses, with the symmetric extensions to negative R highlighting mathematical completeness, albeit physical applicability is restricted to R > 0. Overall, in this Lorentz-breaking cavity setup, $l_b$ parameterically modulates thermodynamic scales by redshifting and localising energy, essentially reformulating equilibrium conditions, heat capacities, and ensemble viability for black hole analogues, allowing sharper theoretical windows for probing asymmetry-driven gravitational impacts and potential empirical correlations with compact object observations.

\subsection{Free Energy}
% Free Energy
\begin{equation}
F=E-T(R)S
\end{equation}
so that
\begin{align}
F&=-\frac{(1+l_{b})\Lambda R^{2}}
{\sqrt{-M-(1+l_{b})\Lambda R^{2}}}
-\;T(R)\,S \notag\\[0.5em]
&=-\frac{(1+l_{b})\Lambda\bigl((1+l_{b})\Lambda r_{+}^{2}+4R^{2}\bigr)}
{4\sqrt{-M-(1+l_{b})\Lambda R^{2}}}\,.
\end{align}
\begin{figure}[h!]
\centering
\includegraphics[width=0.45\textwidth]{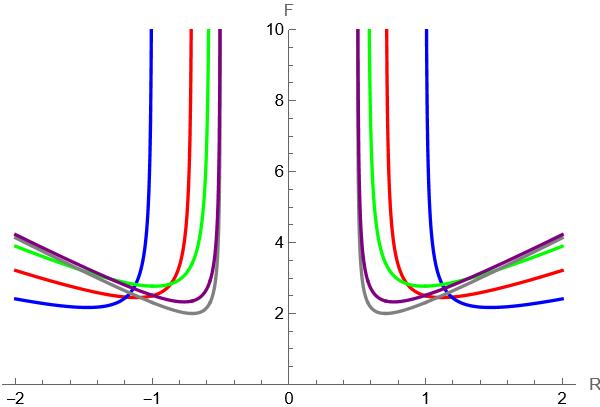}
\caption{Free Energy in York Cavity}
\label{fig:entropy-correction1}
\end{figure}

In the York cavity formalism, the behaviour of the thermodynamical potentials for the $2+1$ dimensional BTZ black hole in bumblebee gravity exhibits marked corrections due to the Lorentz-violating parameter $l_{b}$ and the cavity radius $r$. The graph plots six branches corresponding to different $l_{b}$ and $r$ options. For the first three curves, the trend of $l_{b}$ with values $0, -0.5, 0.5$ (plotted in red, blue, and green, respectively) indicates the influence of the background of the bumblebee on the equilibrium configuration of the black hole in the cavity. A positive $l_{b}$ lifts the Potential to greater values, i.e., the Lorentz-violating effects enhance the thermodynamical stability. In contrast, a negative $l_{b}$ lowers the curves, describing a suppression of stability and an increase in the tendency for phase transition.

For the remaining three curves, where $l_{b}$ is kept constant and the cavity parameter $r$ changes with the values $0, -0.5, 0.5$ (in grey, purple, and black), the effect of the cavity geometry can be observed. As $r$ increases, free energy increases, and the position of extrema is shifted, indicating that the size of the cavity directly determines the equilibrium states allowed. Conversely, an unfavourable change in $r$ lowers the thermodynamic Potential, which is correlated with larger boundary effects and possible reduced stability. The interaction between both parameters highlights that the York formalism preserves the intrinsic black hole thermodynamics and the Lorentz violation and boundary condition effects, thereby enhancing the structure of possible stable and metastable configurations compared to the standard BTZ case.

\subsection{Stability in York cavity}
% Specific Heat
\begin{equation}
C=\frac{dE}{dT}
=\frac{2\pi\bigl(2M+(1+l_{b})\Lambda R^{2}\bigr)}
{(1+l_{b})\Lambda r_{+}}\,.
\end{equation}

\begin{figure}[h!]
\centering
\includegraphics[width=0.45\textwidth]{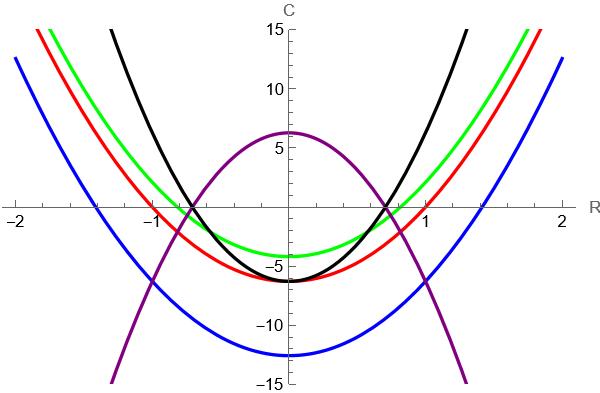}
\caption{Specific Heat in York cavity formulism }
\label{fig:entropy-correction2}
\end{figure}

In the York cavity formalism, the specific heat of $2+1$ dimensional BTZ black hole in bumblebee gravity is a crucial probe for the local thermodynamical stability of the system. The six curves correspond to the same interval of variation of the Lorentz-violating parameter $l_{b}$ and the cavity parameter $r$ as in the above discussion. In the first three curves, where $l_{b}$ is varied as $0, -0.5, 0.5$ (red, blue, and green curves respectively), the bumblebee field effect shifts the value and sign of the specific heat. A positive $l_{b}$ extends areas of positive specific heat, translating to an increase in the amount of thermodynamically stable phases. In contrast, a negative $l_{b}$ would extend areas of negative specific heat, meaning higher instabilities and even phase transition possibilities. 

For the remaining three curves, where the cavity parameter $r$ is swept as $0, -0.5, 0.5$ (grey, purple, and black curves), boundary condition influence comes into play. Rising $r$ is apt to push the curve upwards into positive specific heat regions and thus prefers stability inside the cavity—conversely, negative values of $r$ force the system towards more unstable configurations with negative specific heat. The presence of a sign change in the particular heat is tantamount to critical points where the system undergoes transitions from the stable to unstable branch. The results confirm that the Lorentz-breaking background and the cavity shape control stability and produce a richer phase structure when their interaction is turned on compared to the conventional BTZ black hole. The York cavity approach, therefore, opposes the way boundary effects simultaneously with bumblebee corrections manage the thermodynamical behaviour of the system.

\section{Final Remarks}

In this work, we have investigated the thermodynamics of $2+1$ dimensional BTZ black holes in the framework of Bumblebee gravity, incorporating both Barrow entropy corrections and the York cavity formalism. Our analysis can be divided into two complementary aspects: (i) modifications arising from Lorentz-symmetry violation and fractal Entropy in asymptotically AdS spacetime, and (ii) quasilocal thermodynamics induced by finite cavity boundary conditions.  

\subsection{Bumblebee Gravity and Barrow Entropy Effects}

The introduction of the Lorentz-violating Bumblebee parameter $l_b$ directly modifies the black hole horizon structure, resulting in rescaling of the thermodynamic quantities such as Entropy, internal energy, Helmholtz and Gibbs free energies, and Pressure. Specifically, a positive $l_b$ enhances the Entropy and shifts all thermodynamic potentials upwards, whereas a negative $l_b$ suppresses these quantities. The linear dependence of classical Entropy $S \propto r_{+}$ is preserved, indicating that Bumblebee modifications primarily rescale the thermodynamic behaviour rather than changing the functional form.  

When Barrow entropy is incorporated through the fractal exponent $\Delta$, we observe a fundamentally different behaviour: the Entropy, free energies, and internal energy exhibit superlinear or nonlinear dependence on the horizon radius $r_{+}$. This reflects the increased number of microstates associated with a fractalized horizon geometry. For larger $\Delta$, the Helmholtz and Gibbs free energies become more convex, the internal energy declines more steeply, and the pressure exhibits amplified divergence near small $r_{+}$. These observations imply that Barrow entropy corrections introduce strong nonlinearities in the thermodynamic potentials, which can significantly alter stability criteria, phase transitions, and heat capacity profiles. Therefore, the combined effects of $l_b$ and $\Delta$ provide a rich two-parameter deformation of classical BTZ thermodynamics, with one acting as a linear scale factor and the other as a nonlinear, fractal-induced modification.  

\subsection{York Cavity Formalism and Local Thermodynamics}

We employed the York cavity formalism to construct a well-defined canonical ensemble, placing the black hole inside a finite spherical cavity of radius $R$ with fixed boundary temperature $T(R)$. According to Tolman's law, the local temperature, quasilocal energy, and free energies are redshifted within this framework. The Lorentz-violating parameter $l_b$ modulates both the slope and amplitude of the V-shaped local temperature $T(R)$, with positive $l_b$ steepening the gradient and enhancing thermal response, while negative $l_b$ mitigates it.  

Similarly, the Brown–York quasilocal energy $E_{\text{BY}}$ exhibits symmetric parabolic growth as a function of $R$, with $l_b$ controlling the curvature of the parabola. Positive $l_b$ amplifies energy localisation at the boundary, whereas negative $l_b$ spreads the energy more uniformly, indicating that Lorentz violation regulates energy distribution within the cavity. The free energy and specific heat calculated in the York formalism demonstrate that the cavity size $R$ stabilises: larger $R$ favours positive specific heat and thermodynamic stability, whereas minor or negative shifts in $R$ can produce unstable branches. This highlights the dual role of $l_b$ and the cavity radius as regulators of equilibrium and phase structure.  

\subsection{Synthesis of Results and Outlook}

The initial changes start with the BTZ metric brought into bumblebee gravity form, resulting in an effective cosmological constant and changed horizon characteristics. The Hawking temperature, area, and volume are obtained, with the result that $l_b$ rescales the effective gravitational interactions without essentially changing the qualitative structure in the absence of Barrow corrections. The entropy retains the standard Bekenstein-Hawking structure but is rescaled by $\sqrt{1 + l_b}$ and grows linearly with horizon radius $r_+$. Adding Barrow entropy converts this to a superlinear dependence $S_B \propto (r_+)^{1 + \Delta/2}$, indicating quantum-gravitational fractalization of the horizon. Graphical calculations validate that $l_b$ linearly adjusts the slope, whereas $\Delta$ adds convexity, augmenting entropy for larger horizons and implying more abundant microstate counting.

Thermodynamic potentials like Helmholtz free energy $F$, pressure $P$, internal energy $U$, enthalpy $H$, and Gibbs free energy $G$ are calculated both with and without Barrow corrections. In the traditional example, these potentials exhibit quadratic or inverse dependences on $r_+$, modulated by $l_b$ through scaling factors: positive $l_b$ raises magnitudes, and negative ones suppress them. Barrow corrections cause nonlinear expansions, resulting in shallower profiles, sign changes in particular potentials (e.g., $H$ and $G$), and asymmetric behaviors, which suggest potential metastable phases or modified stability thresholds. The heat capacity $C$ is still positive and monotonically increasing with $r_+$, but Barrow terms enhance the increase, especially for positive $l_b$, indicating greater thermal stability and responsiveness.

Moving to York's cavity formalism, the Tolman-redshifted boundary temperature $T(R)$ has a V-shaped, $R=0$-symmetric profile, with slopes determined by $l_b$ and a further radial parameter $r$ (potentially a horizon-related term). Negative $l_b$ or $r$ tempers the increase, encouraging wider stability, whereas positive values steepen it, potentially generating disequilibria. The Brown-York quasilocal energy $E_{BY}$ displays parabolic increase, once more scaled by $l_b$, describing boundary-contained energy densities dependent on Lorentz violations. The cavity free energy $F$ and specific heat $C$ reproduce these tendencies, with $l_b$ and $r$ dictating minima, sign reversals, and stability regions. Positive specific heat regions enlarge upon positive $l_b$ or $r$, whereas negatives create instabilities, increasing the phase structure.

Together, these results emphasize the combined actions of Lorentz-breaking bumblebee gravity and quantum-motivated Barrow entropy on BTZ black hole thermodynamics. Linear rescalings from bumblebee corrections, reminiscent of effective couplings, are contrasted with nonlinear, power-law variations through Barrow deformations that simulate higher-dimensional or quantum influences and potentially create critical behavior missing in standard (2+1)D configurations. The York cavity assures physical finiteness, exhibiting boundary-dependent stability and phase transitions that can guide holographic projections in AdS/CFT.

This research takes our knowledge of modified gravity theories in lower dimensions forward, providing a testable model for quantum gravity phenomenology. The resultant changes in thermodynamic profiles are indicative of observable signatures in analog systems or gravitational wave signatures from compact objects. Extensions in the future may include rotations, charges, or other corrections (e.g., logarithmic entropy corrections), global stability through Gibbs ensembles, or probing holographic duals to boundary CFTs under Lorentz violations. Finally, these findings cross classical relativity, quantum corrections, and symmetry-breaking paradigms to open the possibility of unified formulations of gravitational thermodynamics in exotic regimes.

\end{document}